\documentclass[submission, Phys]{SciPost}

\usepackage[T1]{fontenc}
\usepackage{booktabs}
\usepackage{xspace}
\usepackage{hyperref}
\usepackage{subcaption}
\usepackage{fontawesome5}
\usepackage{orcidlink}

\hypersetup{
    colorlinks,
    linkcolor={red!50!black},
    citecolor={blue!50!black},
    urlcolor={blue!80!black}
}

\usepackage[bitstream-charter]{mathdesign}
\newcommand{\DL}{\texttt{DL1}\xspace}
\newcommand{\GN}{\texttt{GN2}\xspace}
\newcommand{\Opus}{\texttt{Opus\,5.5}\xspace}

\begin{document}
\begin{center}{\Large \textbf{
Ceci n'est pas un tagger\\
Can a large language model flavour-tag a jet?
}}\end{center}

\begin{center}
Andrea Coccaro\,\orcidlink{0000-0003-2368-4559}\textsuperscript{1},
Andrea Milici\,\orcidlink{0009-0002-4434-7404}\textsuperscript{1,2},
Lucio Derin\,\orcidlink{0000-0001-6507-114X}\textsuperscript{1,2}
\end{center}

\begin{center}
{\bf 1} INFN, Sezione di Genova, 16146 Genova, Italy
\\
{\bf 2} Department of Physics, Università di Genova, 16146 Genova, Italy
\end{center}

\section*{Abstract}
Large language models (LLMs) are increasingly capable of reasoning about scientific problems, effectively mastering and synthesizing entire fields of research and providing, for the first time, access to a distilled and queryable form of scientific knowledge.
As the integration of LLMs rapidly spreads across the large collaborations of high-energy physics, aiming to rethink and optimize analysis workflows, it remains unclear whether the knowledge acquired during pretraining can support quantitative inference on physical data.
Moving beyond the applications currently pursued by the community, we pose a simple question: can an LLM classify the flavour of a hadronic jet? An LLM is therefore asked to assign flavour probabilities ($b$, $c$, light, $\tau$) to jets from a public ATLAS flavour-tagging sample, using only reconstructed constituents. Under a small set of constraints, the model is not allowed to write code or implement conventional algorithms and must instead perform the inference directly from its pretrained knowledge and the available inputs.
We report two complementary results and make publicly available, through a GitHub repository~\href{https://github.com/FTAGwLLM/ceci-nest-pas-un-tagger}{\faGithub}, all the code required to reproduce the experiment and obtain the quantitative results presented in this work. 
First, the model achieves non-trivial discrimination between flavours, which we compare with recent ATLAS flavour-tagging algorithms. Second, the model accompanies each prediction with a natural language rationale. The physical concepts invoked in these rationales correlate with the truth flavour and reproduce the features traditionally associated with each jet category. Flavour tagging thus provides a laboratory for broader questions: how can we actively leverage the corpus of scientific knowledge accumulated over decades of research? Could LLMs serve as a general source of scientific context for task-specific algorithms?

\newpage
\section{Introduction}
\label{sec:intro}
Hadronic jets, collimated sprays of particles originating from quarks or gluons, originate abundantly from the proton--proton collisions at the Large Hadron Collider (LHC)~\cite{ref:LHC}.
Flavour tagging, the process of identifying the flavour of the parton initiating a jet, is a critical component of the physics programme of the ATLAS and CMS experiments~\cite{ref:ATLAS-Paper,ref:CMS-Paper}.
Jets originating from $b$- and $c$-quarks are experimentally distinguishable from light-flavour ($u,d,s$) quark or gluon jets owing to the characteristic properties and decay patterns of the heavy-flavour hadrons they contain~\cite{Guest:2016iqz}.
Hadronically decaying $\tau$ leptons also exhibit a distinct set of signatures, typically producing narrow, low-multiplicity jets characterized by one or three charged prongs.

Over the past decade flavour-tagging algorithms have tremendously evolved~\cite{Mondal:2024nsa}. From extracting physics-inspired observables and constructing a classifier algorithm with deep neural network architectures~\cite{ATLAS:2022qxm,Bols:2020bkb} to an end-to-end transformer-based architecture with auxiliary training objectives to enhance both interpretability and overall classification performance~\cite{Qu:2022mxj,ATLAS:2025dkv}.
The latest incarnations of flavour-tagging algorithms are based on the attention mechanism, which has revolutionized machine learning by enabling models to capture complex relationships within structured and sequential data~\cite{Vaswani:2017lxt}.
Each new approach has brought substantial gains in classification performance when evaluated on simulated samples, with only mild degradation observed when these gains are translated to data, as assessed through dedicated calibration analyses~\cite{ATLAS:2026bvi}.

Flavour tagging follows the supervised machine-learning paradigm. A parametric function $f_\theta$ can be constructed to map the reconstructed observables $x$ of a jet to flavour probabilities,
\begin{equation}\label{eq:1}
x \;\longrightarrow\; f_\theta(x) \;\longrightarrow\;\mathbf{p}_i = \big(P(b \mid x_i),\, P(c \mid x_i),\, P(u \mid x_i),\, P(\tau \mid x_i)\big), \qquad \sum_k p_{i,k} = 1
\end{equation}
with the parameters $\theta$ determined by minimising a cross-entropy loss on simulated samples in which the truth flavour is known.
The effectiveness of this approach therefore relies not only on the expressive power of the model but also on the availability and fidelity of the simulated data used for training. This reliance on simulation is particularly effective in HEP, where detailed detector simulations provide an accurate description of the relevant physics processes and detector response. More recent training strategies have exploited increasingly large simulated datasets, reaching substantially higher statistics than earlier generations of established classification algorithms. This development has been motivated in part by the observation that transformer-based architectures continue to benefit from increasing training statistics, with their performance showing no clear sign of saturation over the range of dataset sizes explored so far~\cite{ATL-SOFT-PUB-2026-002}.

Large language models (LLMs) are neural networks trained to predict the next token on very large corpora that include a substantial fraction of the scientific literature: articles, textbooks, technical notes, and software documentation.
Recent models are further tuned to follow instructions, to reason over multiple steps~\cite{Wei:2022CoT}, and to act as agents that use tools~\cite{Yao:2023} to solve non-trivial tasks.  LLMs have been shown to exhibit emergent capabilities, namely reasoning abilities or other behaviours that are not observed in smaller models and cannot be straightforwardly anticipated by extrapolating their performance~\cite{Wei:2022Emergence}.

In this work, we investigate whether the scientific knowledge acquired by an LLM during training can be used to infer the flavour of a jet directly from its reconstructed observables.
We consider a zero-shot setting, similarly to other recent works in other science domains~\cite{Zero-Shot-1} or HEP applications~\cite{Zero-Shot-2}, in which the LLM performs the classification task without being provided with labelled examples or undergoing task-specific training.
Applying such capabilities to flavour tagging would represent a fundamentally different paradigm from supervised approaches, in which the association between reconstructed observables and jet flavour is learned from labelled training samples.
An LLM, by contrast, may already encode aspects of the underlying physics in its parameters through its exposure to a vast scientific literature during training.
Such an emergent capability would therefore amount to transferring scientific knowledge acquired during general-purpose training to a specialised classification task for which the model was never explicitly trained on.
Equation~\ref{eq:1} would then become: 
\begin{equation}\label{eq:task}
x_i \;\longrightarrow\; \mathbf{p}_i = \big(P(b \mid x_i, \varphi),\, P(c \mid x_i,\varphi),\, P(u \mid x_i,\varphi),\, P(\tau \mid x_i, \varphi)\big), \qquad \sum_k p_{i,k} = 1 \, .
\end{equation}
Here $\varphi$ represents the physical information encoded in the LLM model weights.
The study presented in this work is organised around three questions: does the LLM achieve a non-trivial discrimination between jet flavours? Which physical features does the model invoke to justify its predictions? Do these features correspond to the physics knowledge traditionally associated with flavour tagging?

Section~\ref{sec:tag} describes the experimental setup and presents the resulting performance. Section~\ref{sec:explain} discusses the physics features, expressed in textual form, that drive the predictions. The final section overviews the results, their limitations, and discusses future prospects.

\section{Can an LLM tag a jet?}
\label{sec:tag}

The experiment consists on asking an LLM to flavour-tag a jet by quantitatively returning probabilities for the flavour hypothesis given the jet reconstructed observables and under the explicit zero-shot constraint.
The LLM is explicitly not allowed to construct any algorithmic approach and receives neither truth-level flavour information nor scores of existing algorithms trained within the collaborations.
In this way, the LLM reasoning capabilities for flavour-jet physics are tested and not its ability to construct from first principles a software for classification purposes.
The adopted LLM is Claude by Anthropic and the actual model is \Opus, one of the best released models at the time of writing~\cite{Anthropic:Opus55}.

The input dataset is an official sample released publicly by the ATLAS collaboration~\cite{ATLAS:JetSet}. It contains 50 million events from simulated top quark pair production at a centre-of-mass energy of 13.6~TeV. It contains structured event-level, jet-level, track-level, and truth-hadron information compatible with the purpose of flavour-tagging algorithm development. 
The actual dataset being used in this experiment is a subset of 10k~jets where reconstructed jet and constituent-level observables are preserved, while truth and flavour information is completely removed.

For each jet under scrutiny, \Opus is required to infer the four probabilities, constrained to sum to unity.
The task is to use the model's internal reasoning capabilities when performing the flavour inference by integrating the available observables, the learned physics knowledge, and the relationships between different reconstructed quantities in the model's latent space.
The task also explicitly forbids to externalise or replace the reasoning with a hand-crafted algorithm, explicit scoring formula, or programmatic procedure.
In addition, the experiment is encapsulated in a particular directory where no additional information is provided and it is run isolated from both the rest of the filesystem and the web.
A whole list of additional restrictions is given as instructions and can be consulted in the GitHub repository~\href{https://github.com/FTAGwLLM/ceci-nest-pas-un-tagger}{\faGithub} accompanying this paper.
All these guardrails are to ensure that the scored probabilities are the result of the model-level inference in latent space given the provided reconstructed data. 

\Opus is deployed within an agentic framework that enables it to autonomously plan and execute a sequence of actions.
 Upon receiving the task specification, the main agent chose to inspect a pilot sample of approximately 40 jets to familiarise itself with the data structure and the meaning of the variables.
It subsequently distributed the remaining jets among different sub-agents, each assigned to process batches of 60 jets.
Each sub-agent received only the task specification and its assigned jets, without any classification criteria or thresholds derived from the pilot sample.
 This autonomous organisation of the workflow is itself a noteworthy aspect of the experiment, although minor variations in the procedure were observed during the initial trial runs.

The full inference procedure required 4 hours and 28 minutes of processing time, increasing to 24 hours and 36 minutes of elapsed time when including interruptions imposed by usage limits and the associated reset periods. By comparison, \GN processes a single jet in $\mathcal{O}(\mathrm{ms})$ on a typical CPU node. This substantial difference in computational cost reflects the fundamentally different nature of the two approaches. While dedicated flavour-tagging algorithms perform a single forward pass through a neural network, the LLM-based approach involves generating and processing sequences of tokens, potentially accompanied by multiple reasoning steps and agentic operations.
The resulting latency is therefore not attributable solely to the classification task itself, but also to the fact that LLMs are inherently designed to process textual tokens rather than numerical data.

\begin{figure}[t!]
\centering
\begin{subfigure}[b]{0.49\textwidth}
\includegraphics[width=\textwidth]{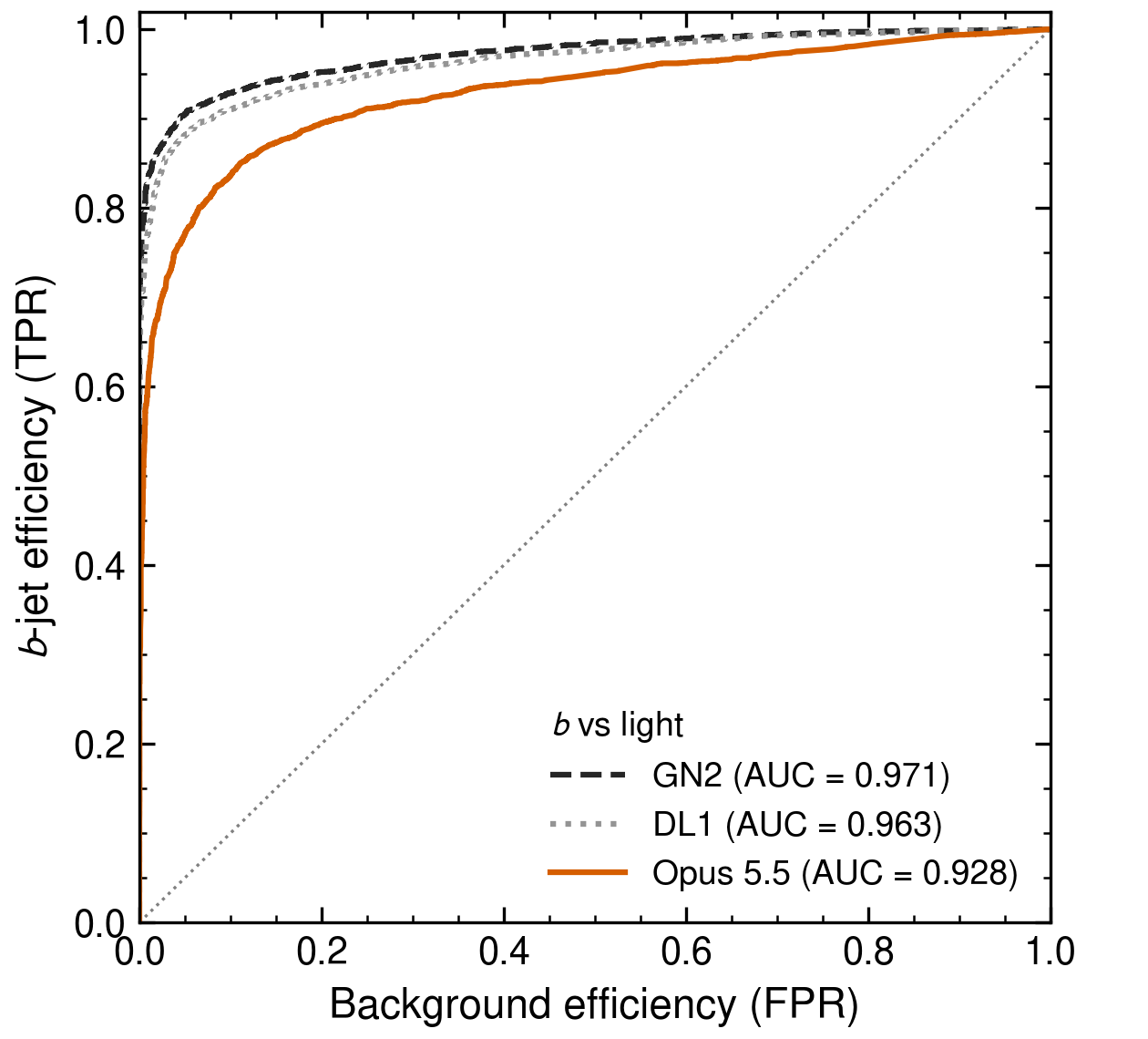}
\caption{}
\end{subfigure}
\begin{subfigure}[b]{0.49\textwidth}
\includegraphics[width=\textwidth]{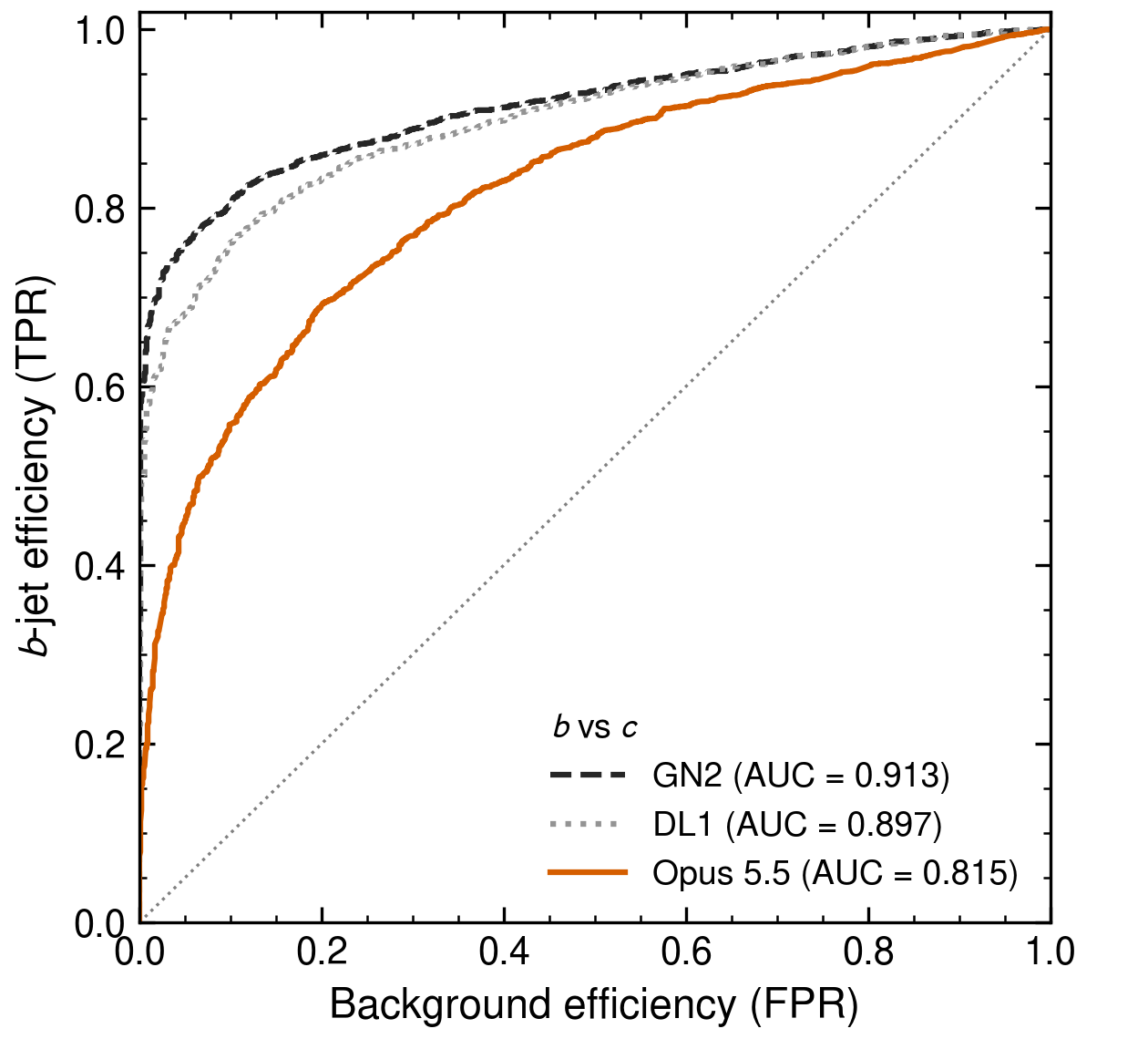}
\caption{}
\end{subfigure}
\caption{ROC curves for (a) $b$- vs light-jets and (b) $b$- versus $c$-jets for the \GN, \DL and the zero-shot inference with \Opus. The corresponding AUC scores are also reported within the figures.}
\label{fig:ROC}
\end{figure}
\begin{figure}[h!]
\centering
\begin{subfigure}[b]{0.49\textwidth}
\includegraphics[width=\textwidth]{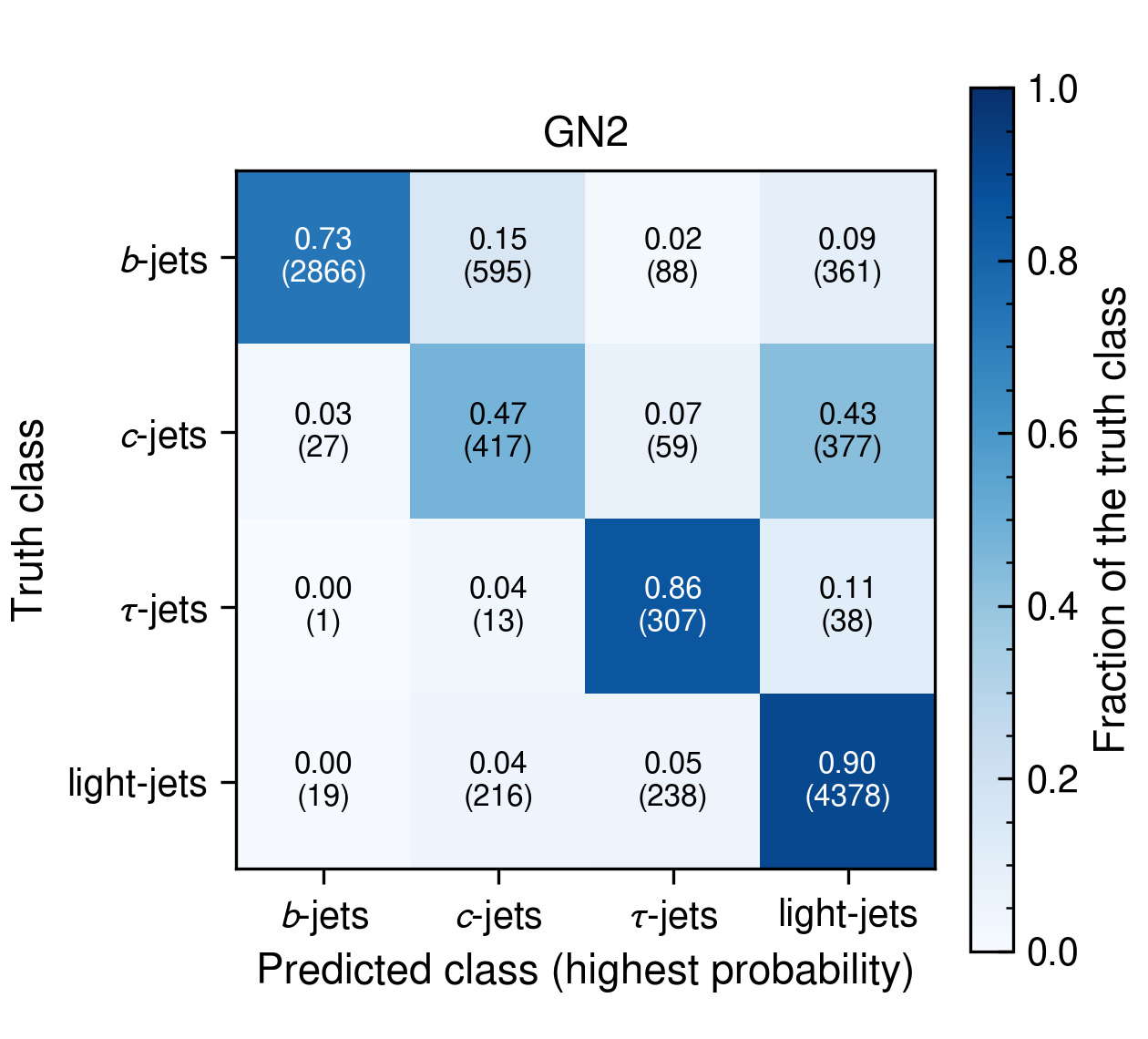}
\caption{}
\end{subfigure}
\begin{subfigure}[b]{0.49\textwidth}
\includegraphics[width=\textwidth]{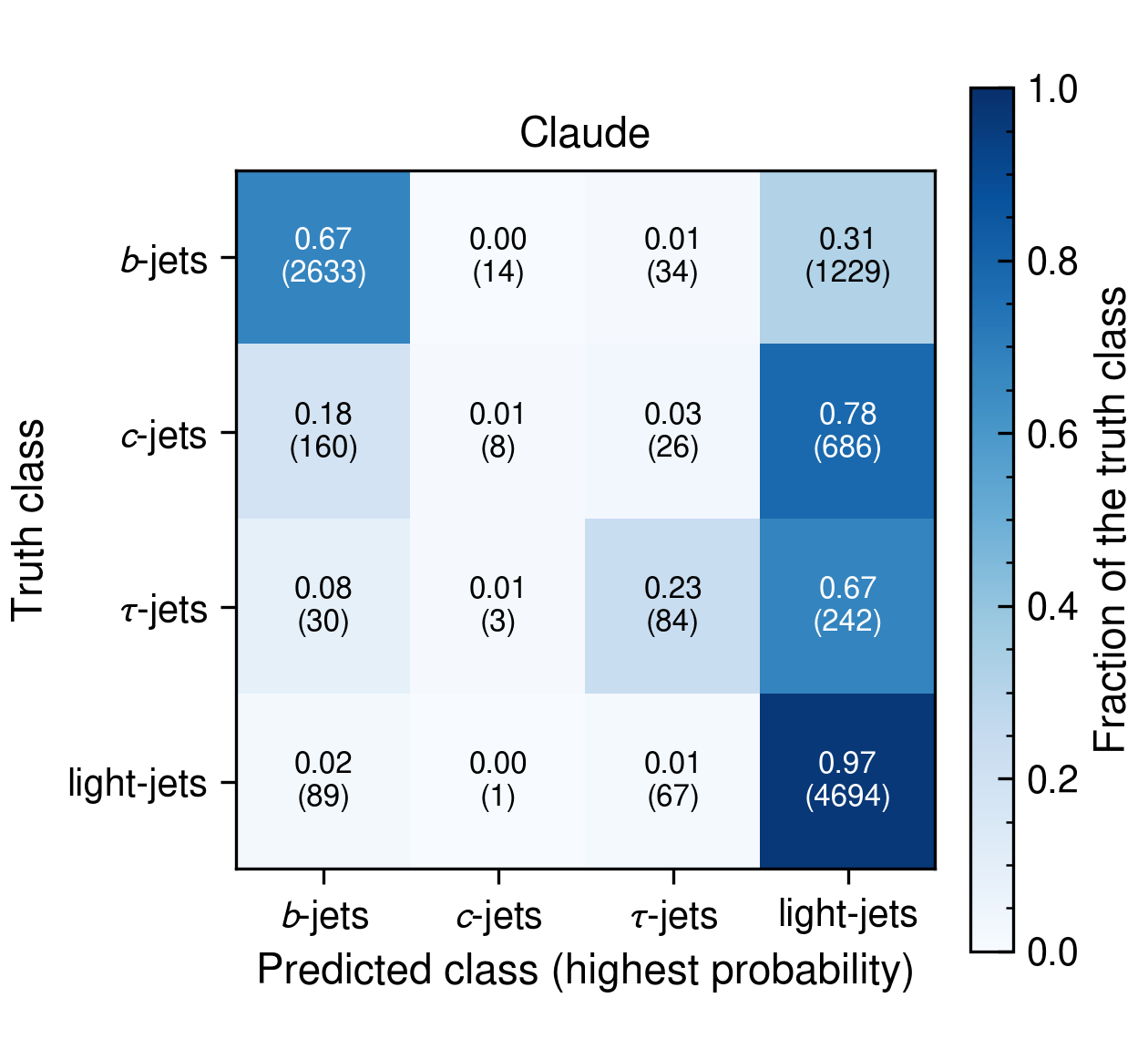}
\caption{}
\end{subfigure}
\caption{Confusion matrices for (a) \GN and (b) \Opus. The matrices illustrate the classification performance in terms of correctly and incorrectly assigned jet flavours, according to the per-jet truth flavour information.}
\label{fig:Confusion}
\end{figure}

After the inference, the LLM prediction is added back to the original files, which include the truth-flavour information and scores of established classification algorithms. 
These algorithms are the \DL and \GN algorithms developed by the ATLAS collaboration.
Discriminants are built from the LLM probabilities in the same form used for the reference taggers,
\begin{equation}\label{eq:Db}
D_b = \ln \frac{p_b}{f_c\, p_c + f_\tau\, p_\tau + (1 - f_c - f_\tau)\, p_u} \,,
\end{equation}
with $f_c = 0.2$ and $f_\tau = 0.01$ for both \GN and \Opus, and $f_c = 0.2$ and $f_\tau = 0$ for \DL.

Figure~\ref{fig:ROC} shows the two ROC curves for the $b$- versus light-jet and $b$- versus $c$-jet classifications.
The AUC scores for the \Opus model are 0.928 and 0.815 for these two cases, to be compared with 0.971 and 0.913 for \GN and with 0.963 and 0.897 for \DL.
The classification performance achieved by the zero-shot LLM substantially exceeds that expected from random guessing, although it remains below that of the dedicated flavour-tagging algorithms developed by the experimental collaborations and specifically trained for this task.
Interestingly, the performance gap is more pronounced in the discrimination between $b$- and $c$-jets, a task that is notoriously challenging from an experimental perspective and, plausibly, also more difficult to describe within the scientific literature. 

Figure~\ref{fig:Confusion} shows the confusion matrices for the jet flavour inference with \GN and \Opus.
These matrices also provide information for the $\tau$-jet class.
One can see that to discriminate between $c$- and $\tau$-jets requires dedicated training and that these two truth classes are easily confused by \Opus and treated as reconstructed light jets.

\begin{figure}[t!]
\centering
\begin{minipage}[b]{0.49\textwidth}
\includegraphics[width=\textwidth]{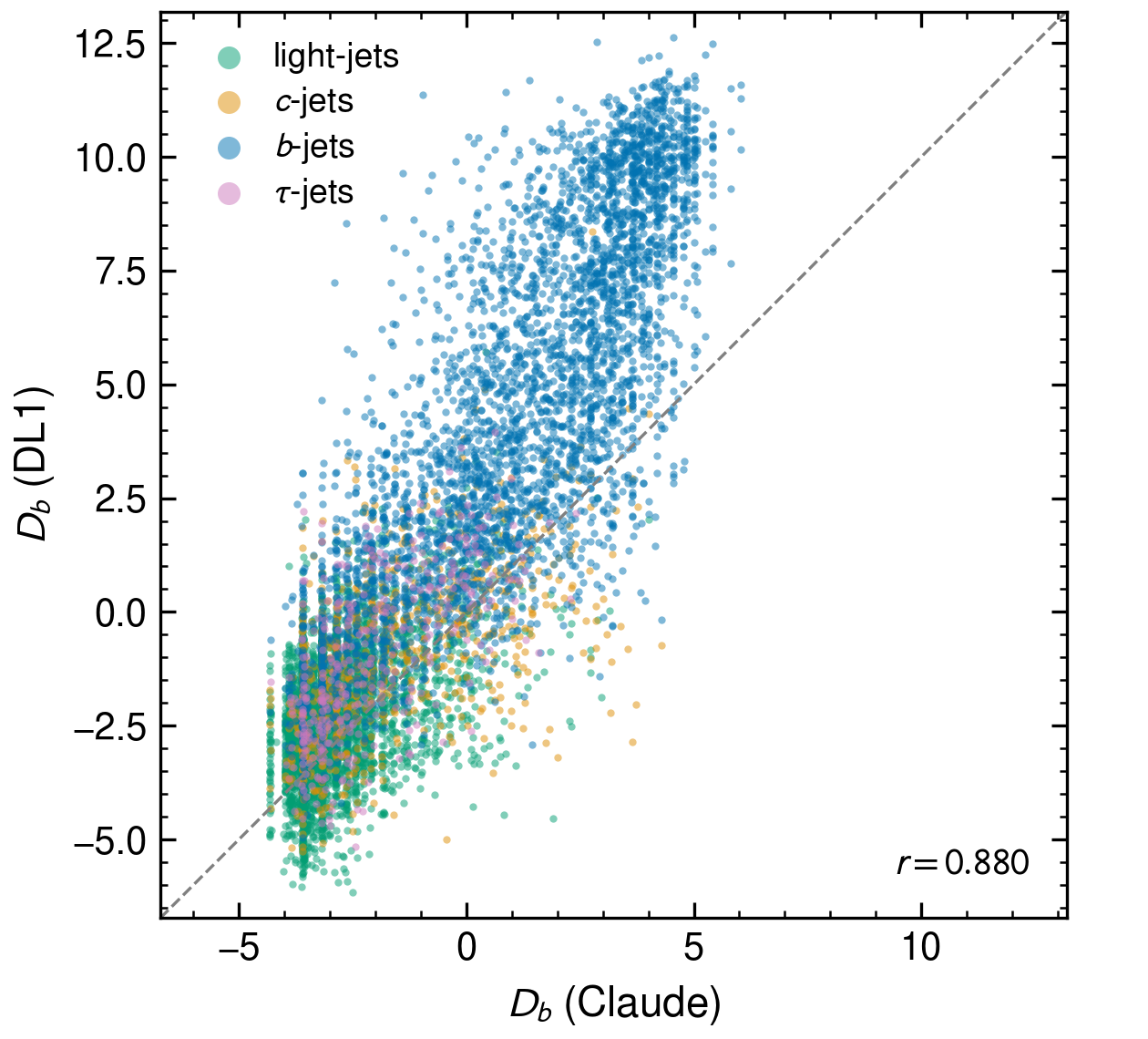}
\end{minipage}
\begin{minipage}[b]{0.49\textwidth}
\includegraphics[width=\textwidth]{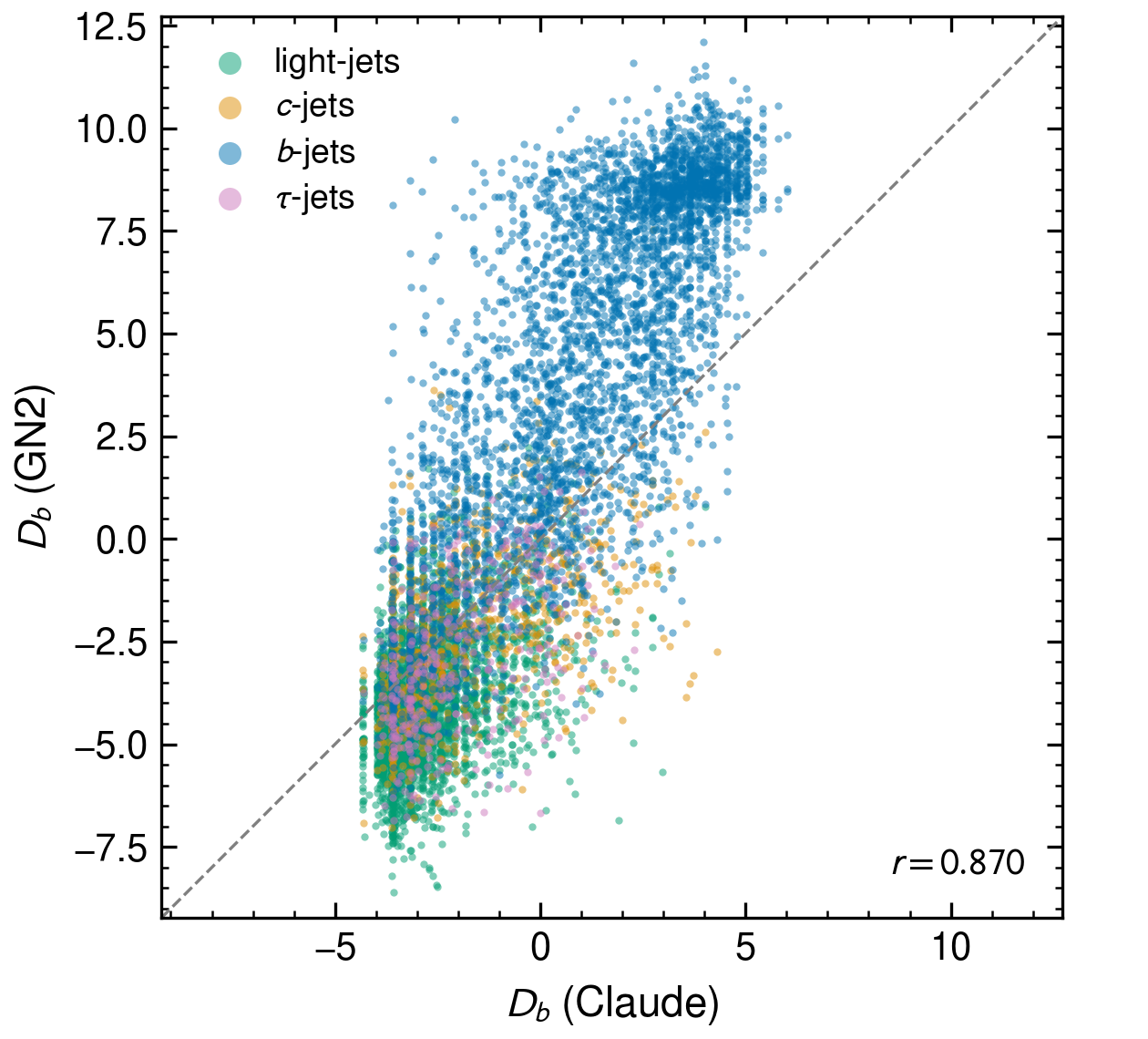}
\end{minipage}
\begin{minipage}[b]{0.49\textwidth}
\includegraphics[width=\textwidth]{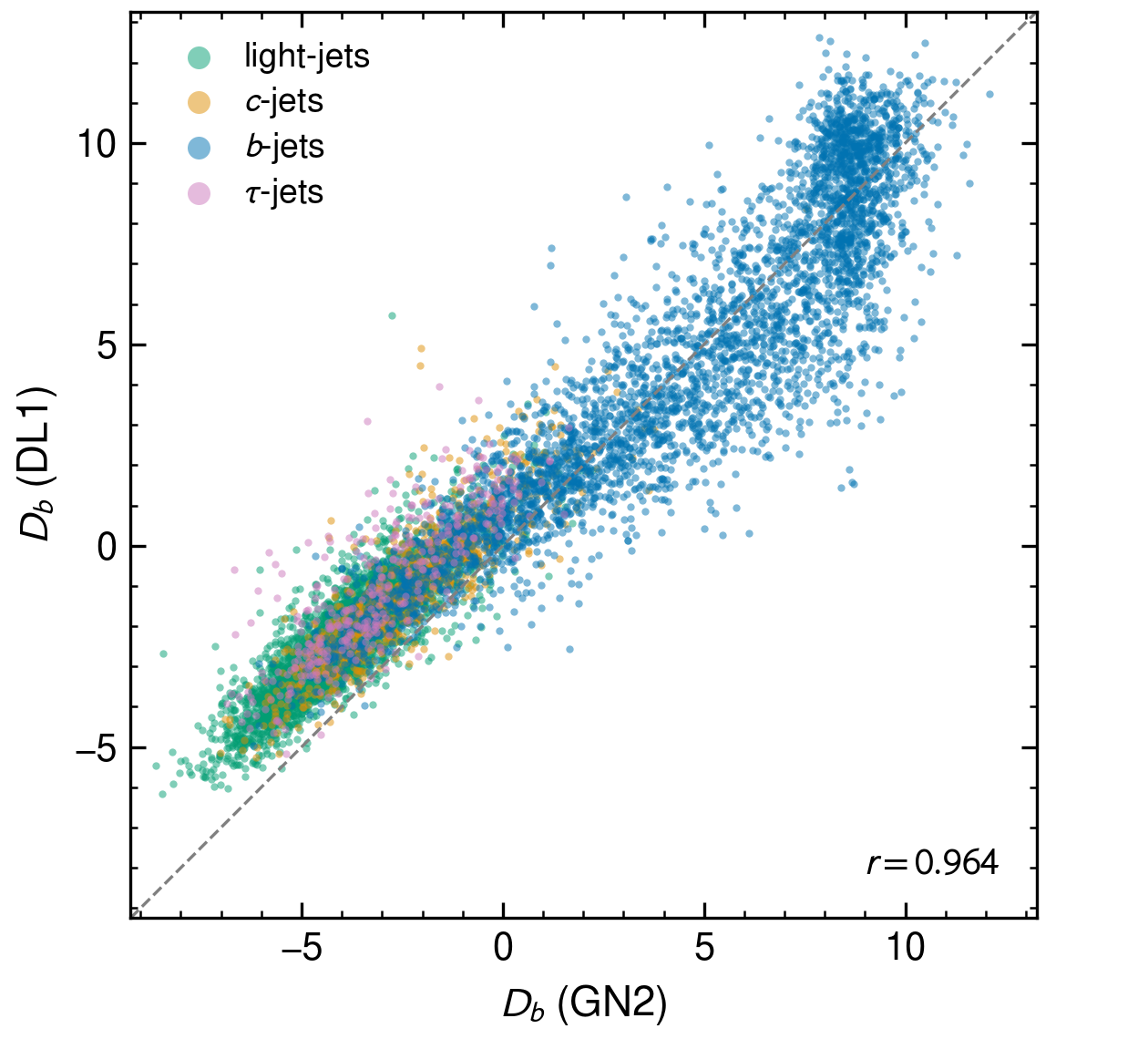}
\end{minipage}
\caption{Two-dimensional distributions of the \(b\)-tagging discriminant \(D_b\) for different tagger pairs, with flavour-separated contributions.}
\label{fig:Discr}
\end{figure}

Finally, the $D_b$ distributions, as defined in Equation~\ref{eq:Db} and their two-dimensional correlations, are reported in Figure~\ref{fig:Discr} with a colour scheme matching the different truth-flavour origin of the jets being inspected. 
The predictions of the \DL and \GN algorithms exhibit a stronger correlation than those of \Opus with either algorithm, as quantified by the Pearson correlation coefficient ($r$). The corresponding $r$ values are 0.964 for \DL and \GN, 0.880 for \Opus and \DL, and 0.870 for \Opus and \GN.
The predictions of \Opus also tend to span a narrower range of values and exhibit a more discrete distribution, potentially indicating a limited expressive capacity of the model for this task.

\section{Emergent explainability}
\label{sec:explain}

The task specification required the predicted probabilities to be accompanied by a brief natural language rationale.
Interestingly, during some initial trial runs, the model spontaneously provided such explanations even when they were not explicitly requested.
The actual inference, as defined in Equation~\ref{eq:task}, can be expanded as 
\begin{equation}
x_i \;\longrightarrow\; (\mathbf{p}_i,\, N_i),
\end{equation}
where $N_i$ is the text associated with jet $i$.

\begin{figure}[t!]
\centering
\begin{subfigure}[b]{0.49\textwidth}
\includegraphics[width=\textwidth]{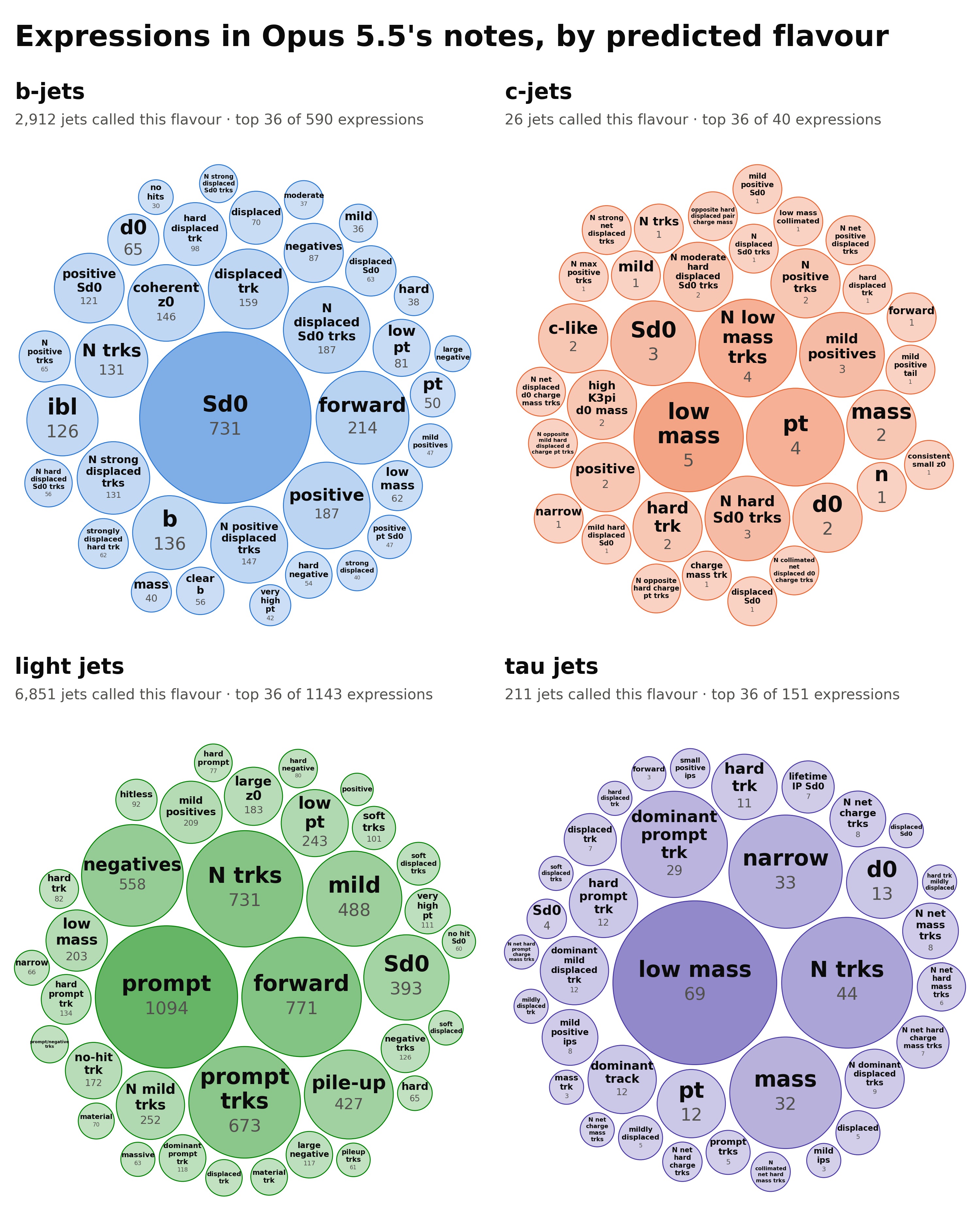}
\caption{}
\label{fig:bubbles_pred}
\end{subfigure}
\begin{subfigure}[b]{0.49\textwidth}
\includegraphics[width=\textwidth]{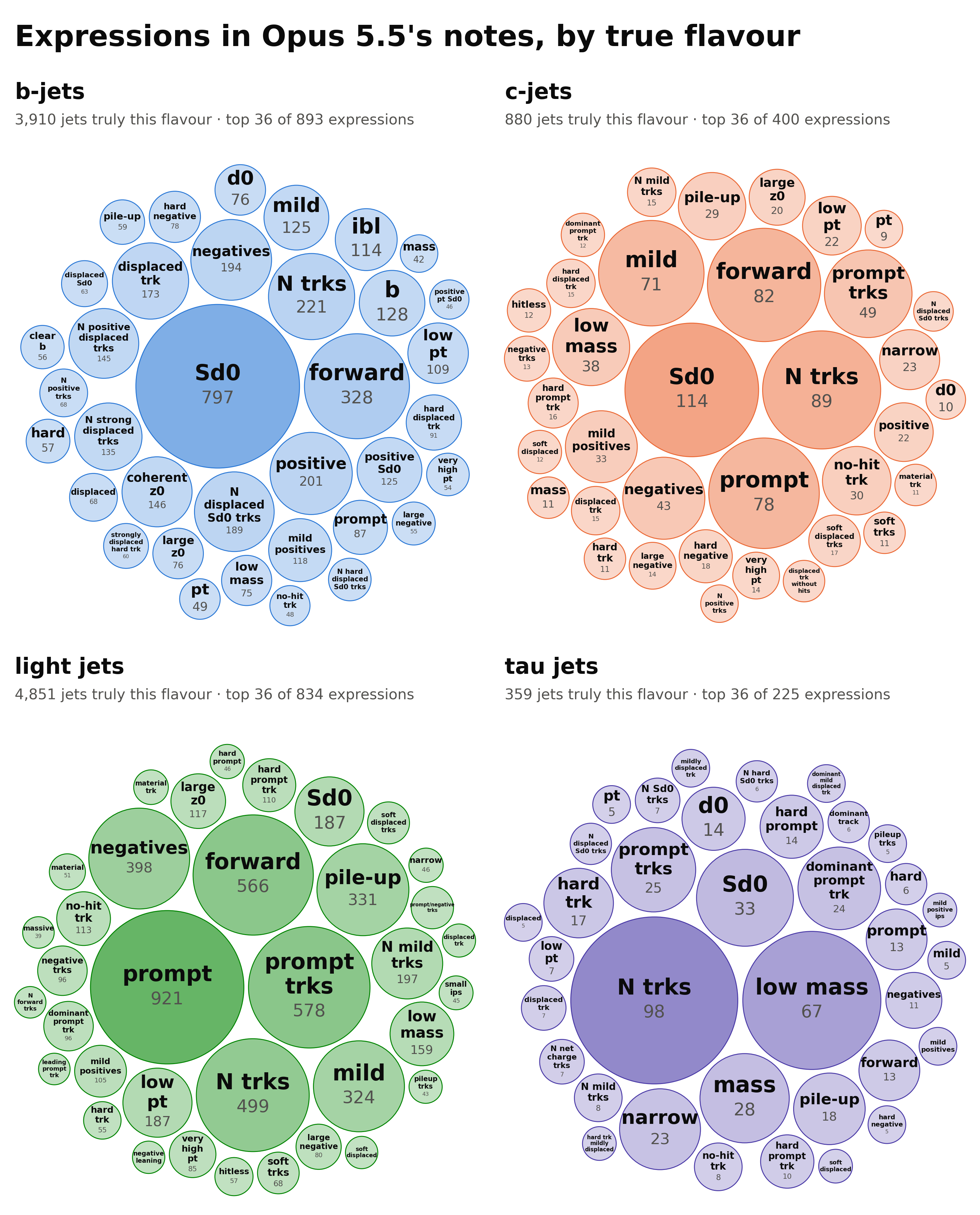}
\caption{}
\label{fig:bubbles_truth}
\end{subfigure}
\caption{Expressions extracted from the text notes generated by \Opus during flavour tagging, counted separately for each flavour and displayed as bubbles. Bubble areas are proportional to the frequency of occurrence, with the corresponding counts indicated below each label. The text notes are divided into clauses each reduced to a set of content words, disregarding word order and cardinal numbers while retaining negations. Each panel displays the 36 most frequent expressions among 1,564 distinct expressions. Jets are grouped according to (a) the flavour assigned the highest probability by \Opus and (b) their truth flavour labels.}
\label{fig:bubbles}
\end{figure}

\begin{figure}[t!]
\centering
\begin{subfigure}[b]{0.49\textwidth}
\includegraphics[width=\textwidth]{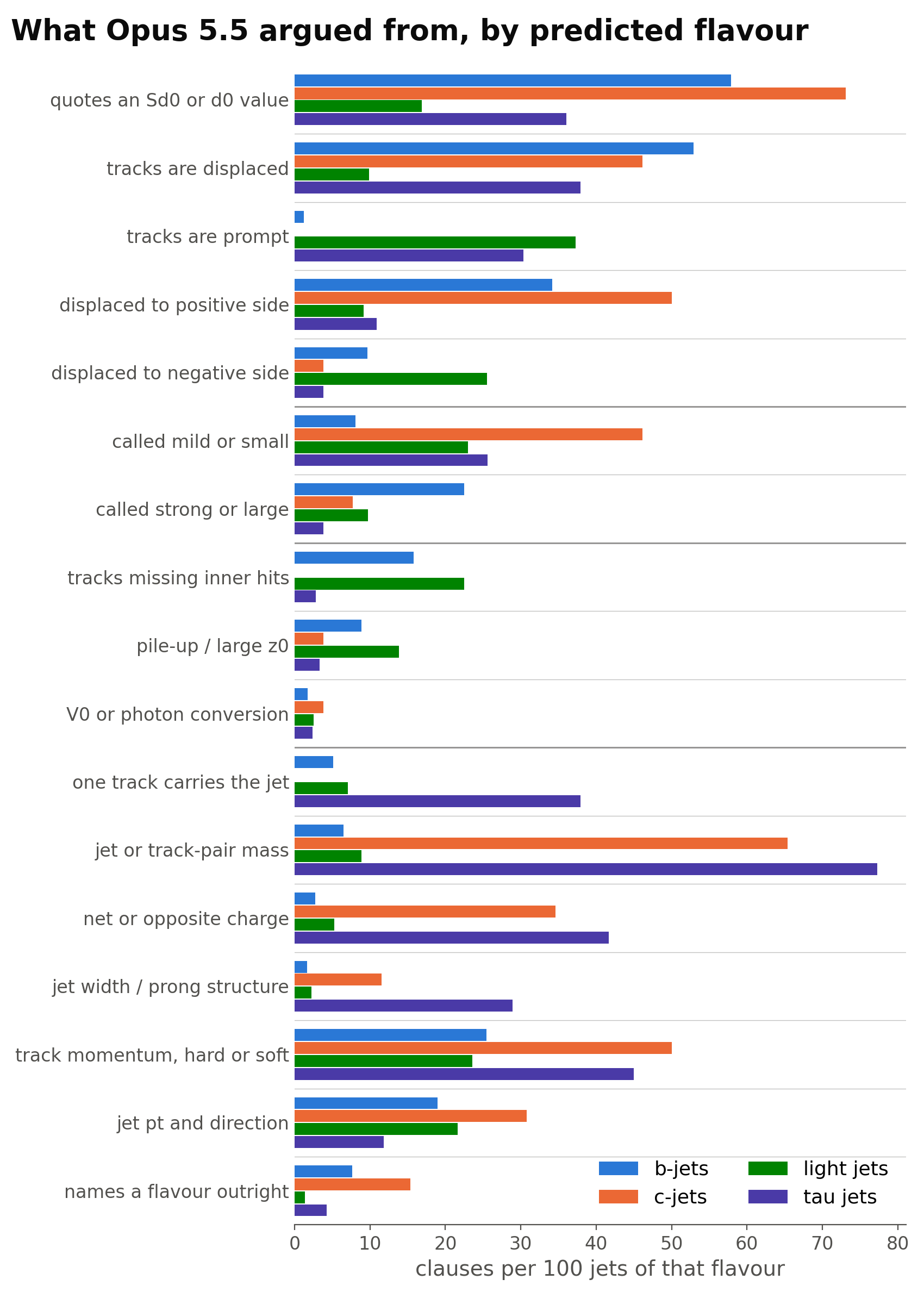}
\caption{}
\label{fig:themes_pred}
\end{subfigure}
\begin{subfigure}[b]{0.49\textwidth}
\includegraphics[width=\textwidth]{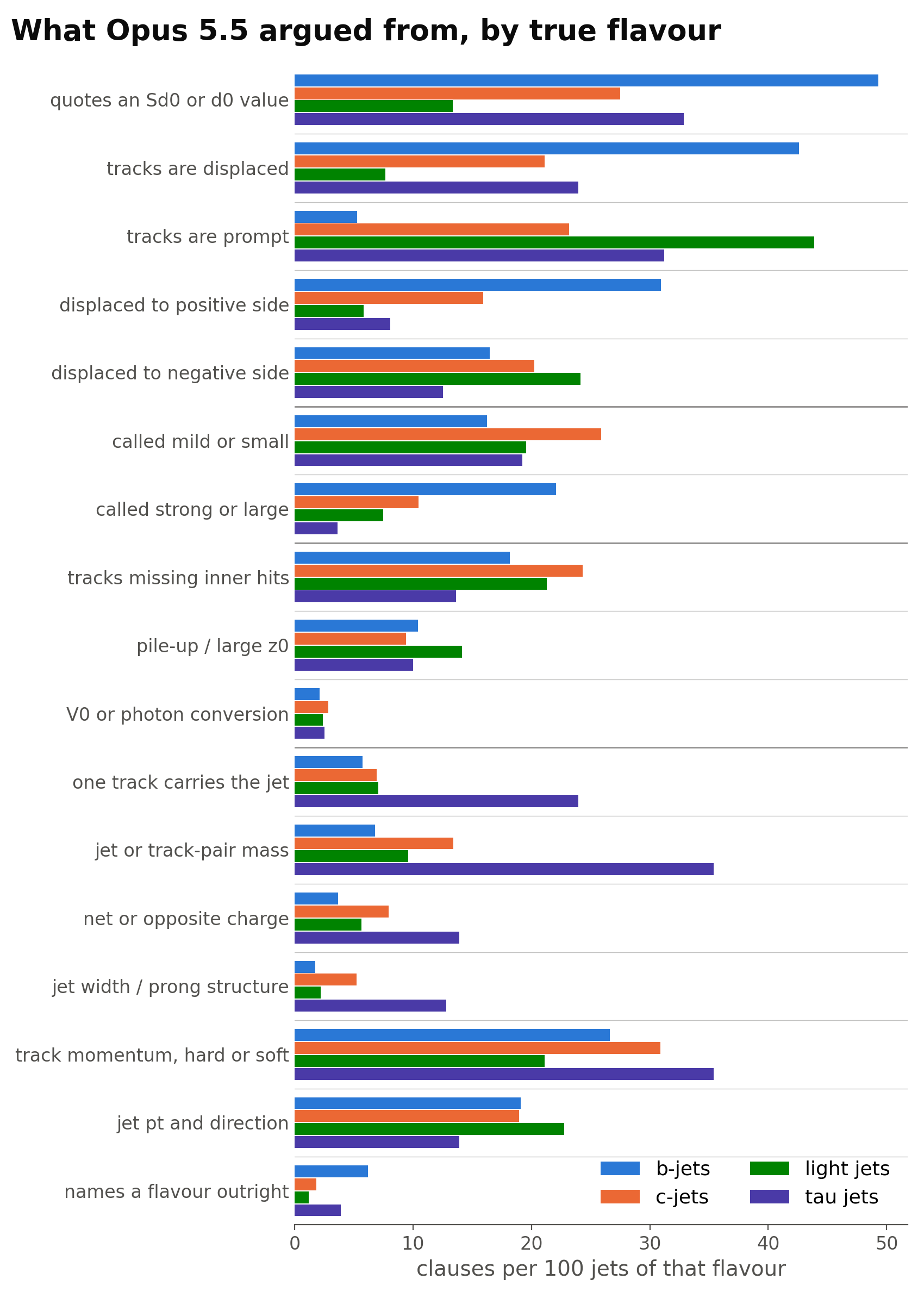}
\caption{}
\label{fig:themes_truth}
\end{subfigure}
\caption{The same clauses grouped by the observable they appeal to, as clauses per 100 jets so the rarer flavours stay comparable. The seventeen themes are a hand-written lexicon matching 95\% of clauses; a clause may support more than one. (a) Grouped by Claude's own call; (b) by truth label, keeping the row order of (a).}
\label{fig:themes}
\end{figure}

The inspection of the text produced when inferring the jet flavours is automated by extracting recurring terms and by displaying the most occurring ones in Figure~\ref{fig:bubbles}.
The terms are displayed in bubbles with their areas proportional to the frequency of occurrence and are grouped according to both the flavour predicted by \Opus and the actual true flavour.
By comparing the most recurring expressions by predicted and by true flavours it is possible to appreciate the closeness of the terms for $b$-jets and light-jets, with common expressions such as the significance of the transverse impact parameter of the associated tracks (Sd0) or the compatibility of the tracks with the primary vertex (prompt), and also how the terms tend to differ for $c$-jets, confirming the poorest classification performance for this class.

We also identify a set of recurring physics-related themes in the jet descriptions.
For each theme, we compute the number of clauses per 100 jets, grouping jets by either their predicted or truth flavour, as shown in Figure~\ref{fig:themes}.
For a given flavour, a theme that occurs more frequently among jets predicted to belong to that flavour than among jets with the corresponding truth label indicates that \Opus places excessive emphasis on that feature when making its predictions.
Conversely, a theme that is more frequent among truth-labelled jets than among predicted jets indicates that the corresponding feature is underutilised.
The most pronounced differences between the two distributions are again observed for $c$-jets.

Some selected themes for $c$-jets are also listed in Table~\ref{tab:themes} with their relative usage in clauses per 100 jets and grouped by true and predicted flavours.
From that table it is possible to note that mass- and charge-related arguments are strongly inflated: a jet or track-pair mass is cited for about 65\% of predicted $c$-jets but only about 13\% of true $c$-jets, and a net or opposite charge for about 35\% versus 8\%. Both cues are physically most characteristic of $\tau$ jets, where they are also over-used (mass 77\% predicted vs.\ 35\% true; charge 42\% vs.\ 14\%), and the model thus appears to treat any mention of a (self-) reconstructed mass or charge as evidence of a heavy-hadron or resonance decay, without checking that the value points to a charm hadron. Positive-side displacement (50\% vs.\ 16\%) and displacement described as ``mild'' or ``small'' (46\% vs.\ 26\%) are inflated in the same way. Since positive-side displacement is in truth most frequent in $b$-jets, using it as a charm argument is a generic heavy-flavour shortcut rather than a charm-specific one, and the fact that it is accompanied by ``mild'' or ``small'' for predicted $c$-jets seems to indicate that \Opus pictures them as $b$-jets with a more moderate SV displacement, which is how they are typically described in the literature. Quoting an $S_{d_0}$ or $d_0$ value is also heavily over-represented in predicted $c$-jets (73\% vs.\ 27\%), which suggests that the presence of a number, rather than its magnitude, is taken as a signature.

Conversely, several themes found in true $c$-jets are essentially absent from predicted $c$-jets, namely prompt tracks, tracks missing inner hits, and jets carried by a single track, so that jets described this way are routed to the light or $\tau$ classes.
A misleading effect appears for displacement to the negative side: its rate is similar in true $b$-, $c$- and light jets but it is concentrated in predicted light jets (26\%, against 10\% for predicted $b$ and 4\% for predicted $c$), and the model seems to read it as a light-jet signature, over-trusting its importance in heavy-flavour rejection.
Finally, some cues are used appropriately or carry little information.
Describing displacement as ``strong'' or ``large'' is the most $b$-specific theme, and its rate is nearly unchanged between true and predicted $b$-jets (about 22\%), whereas pile-up, large $z_0$, $V^0$, or photon conversion, jet $p_\mathrm{T}$ and direction, and missing inner hits occur at similar rates across all four flavours and carry little discriminating power. 

\begin{table}[ht!]
\centering
\begin{tabular}{lrrl}
\toprule
Theme & True $c$ & Predicted $c$ & Behaviour \\
\midrule
Jet or track-pair mass          & 13 & 65 & over-weighted \\
Net or opposite charge          &  8 & 35 & over-weighted \\
Quotes $S_{d_0}$ or $d_0$ value & 27 & 73 & over-weighted \\
Displaced to positive side      & 16 & 50 & over-weighted \\
Called mild or small            & 26 & 46 & over-weighted \\
Names a flavour outright        &  2 & 15 & over-weighted \\
Tracks are prompt               & 23 & $\approx$0 & under-used \\
Tracks missing inner hits       & 24 & $\approx$0 & under-used \\
One track carries the jet       &  7 & $\approx$0 & under-used \\
\bottomrule
\end{tabular}
\caption{Frequency of selected physics-related themes per 100 jets, comparing jets with a true $c$-flavour label and those classified as $c$-jets by \Opus.}
\label{tab:themes}
\end{table}

\section{Conclusion}
In this work, we probe the ability of \Opus to classify the flavour of jets, a task of great interest for LHC physics that the model has learned solely through the scientific literature processed during its training.
The model is asked to assign flavour probabilities to jets from a public ATLAS dataset, stripped of all truth information and with only reconstructed quantities retained.

Any outcome of the experiment would have been interesting a priori.
What we observe is that the model achieves non-trivial discrimination between flavours, well beyond random choice but definitely not able to compete with the most refined algorithms specifically trained for this task, which are capable of also resolving the subtle differences that distinguish $b$-jets from $c$-jets.

The model accompanies each prediction with a short natural language description of the jet.
Whether these notes are a genuine representation of its internal reasoning or a convenient justification of the prediction is difficult to establish.
Nevertheless, the physical features invoked are coherent with the jets' truth flavours and reproduce the characteristics verbally associated with each category. 
Indirectly, this simple natural-language tagging exercise offers a unique perspective on our scientific literature, suggesting that the less detailed and distinctive descriptions of $c$-jets compared with those of $b$-jets may be reflected in the model's performance, much as an imbalanced training sample can influence the outcome of supervised learning.

It is clear that this experiment comes with a list of limitations and cannot be considered a novel approach to flavour tagging.
The probabilistic nature of LLMs is what primarily makes their use in any practical application still inconceivable: a trained algorithm such as \GN is a deterministic function of its inputs, whereas the same jet presented twice to an LLM may receive different predictions.
Moreover, although the observed performance is consistent across different sessions and prompt variations, the results remain difficult to disentangle from the choice of model, reasoning effort, and agentic harness.
In this sense, we have observed an event, which is significant in its own right. Further comparisons across different configurations could provide a broader understanding of the emergent behaviours observed in this study.
We therefore encourage readers to reproduce the experiment using the instructions provided in the public repository and to share their findings.

The computational demands and inference time of this approach are orders of magnitude larger than what would be acceptable in a practical application.
In addition, our results show that the model's probability estimates exhibit limited granularity and span a narrower range of values than those of the reference algorithms.
The model in fact tends to output human-rounded probability values, preferring for example 0.800 to 0.823 even when explicitly instructed to provide numerical outputs at a fixed precision.
This is a clear limitation coming from how the model training is optimized for textual data and not for numerical values.
An interesting avenue for addressing these limitations could be offered by newer models, adopting a decision-based approach producing granular probabilities rather than textual tokens. Whether such models can provide better discrimination for flavour tagging remains an open question to be assessed.

While the integration of LLMs into analysis workflows is already underway, the results presented in this work suggest that they may also be used directly to support quantitative inference on physical data.
Exploiting the representation of fundamental physics concepts encoded in these models to guide classically trained algorithms could provide a framework for a new generation of hybrid solutions that combine the best of both worlds.
In this perspective, the initial epoch of training would no longer necessarily correspond to a state of complete ignorance but can rely on a shared base optimized for the specific experiment, field or task.
Whether this will shape the future of conventional classification algorithms remains to be seen.
What seems nowadays clear is that LLMs will become the standard interface to scientific knowledge, with great potential to accelerate discovery.

Ceci n'est pas un tagger. What this study reveals is not a competitive new flavour-tagging algorithm, but a reflection of the scientific knowledge accumulated through decades of work by the flavour-tagging community.

\label{sec:concl}

\end{document}